\documentclass[prl,twocolumn,preprintnumbers,amsmath,amssymb,floats,citeautoscript,nobalancelastpage,superscriptaddress,showpacs,dvipdfmx]{revtex4}

\usepackage[dvipdfmx]{graphicx}
\usepackage[dvipdfmx]{color}
\usepackage{dcolumn}
\usepackage{bm}
\usepackage{times}
\usepackage{txfonts}
\usepackage{color}

\begin{document}

\title{High-pressure synthesis of an unusual antiferromagnetic metal CaCoO$_{3}$ with GdFeO$_{3}$-type perovskite structure}

\author{T.~Osaka}
\affiliation{Department of Applied Physics and Quantum Phase Electronics Center (QPEC), University of Tokyo, Tokyo 113-8656, Japan}
\author{H.~Takahashi}
\affiliation{Department of Applied Physics and Quantum Phase Electronics Center (QPEC), University of Tokyo, Tokyo 113-8656, Japan}
\author{H. Sagayama}
\affiliation{Institute of Materials Structure Science, High Energy Accelerator Research Organization, Tsukuba, Ibaraki 305-0801, Japan}
\author{Y.~Yamasaki}
\affiliation{Department of Applied Physics and Quantum Phase Electronics Center (QPEC), University of Tokyo, Tokyo 113-8656, Japan}
\affiliation{RIKEN Center for Emergent Matter Science (CEMS), Wako 351-0198, Japan}
\author{S.~Ishiwata}
\affiliation{Department of Applied Physics and Quantum Phase Electronics Center (QPEC), University of Tokyo, Tokyo 113-8656, Japan}
\affiliation{PRESTO, Japan Science and Technology Agency, Kawaguchi, Saitama 332-0012, Japan}

\begin{abstract}
The GdFeO$_{3}$-type perovskite CaCoO$_{3}$ has been successfully synthesized by high-pressure oxygen annealing for the oxygen deficient perovskite. A detailed structural analysis based on synchrotron X-ray diffraction data and a thermogravimetric analysis show that the valence of Co is +4 and the sample is free from oxygen deficiency.  This compound shows an antiferromagnetic ordering at 95 K, which has presumably helical spin arrangement, with keeping the incoherent metallic state down to the lowest temperature. This work demonstrates that the Co$^{4+}$-perovskite oxides exhibit a variety of magnetic phases by the band-width control through the lattice distortion.
\end{abstract}


\maketitle
Perovskite-type oxides with late 3$d$-transition-metal ions in an unusually high-valence state exhibit exotic electronic properties. Typical examples are simple cubic perovskites SrFe$^{4+}$O$_{3}$ and SrCo$^{4+}$O$_{3}$, which show metallic conductivity with an unconventional helimagnetic (HM) ordering at low temperatures and a ferromagnetic (FM) ordering at room temperature, respectively.\cite{1,2,3,4} Whereas the perovskites with Fe$^{3+}$ and Co$^{3+}$ are insulating with a $pd$ charge transfer gap,\cite{5} those with Fe$^{4+}$ and Co$^{4+}$ tend to have a finite density of states near the Fermi level, where the contribution from the oxygen 2$p$ orbital is predominant over that from the transition-metal 3$d$ orbital.\cite{6} It has been suggested that the itinerant oxygen hole plays an important role on the metallic conductivity and HM order in SrFeO$_{3}$.\cite{7} On the other hand, CaFeO$_{3}$ shows a metal-insulator transition at 290 K, below which the valence of Fe ions is disproportionated to be +3 and +5.\cite{8,9,10} The transition in the ground state from metallic in SrFeO$_{3}$ to insulating in CaFeO$_{3}$ can be interpreted as a consequence of the variation in the band width through the enhancement of the orthorhombic distortion, which is called a band-width controlled metal-insulator transition.\cite{13}

 The origin of the room-temperature ferromagnetism in the cubic perovskite SrCoO$_{3}$ has been extensively discussed. In perovskite-type cobalt oxides, the spin-state configuration is a primary factor for the electronic properties, which typically depends on the magnitude relation between the crystal field splitting and the Hund's coupling.\cite{23} However, the situation is more complicated in Co$^{4+}$ perovskites due to the strong $pd$ hybridization. From the experimental and theoretical studies, SrCoO$_{3}$ has been proposed to have the intermediate spin state ($e_{\rm g}^{1}t_{\rm 2g}^{4}$; $S=3/2$).\cite{14,15,35} On the other hand, the first-principles calculations by Kune$\check{\rm s}$ $et$ $al$. indicate that the magnetism of SrCoO$_{3}$ arises from the coherent superposition of various spin configurations.\cite{16} 

 The point to be noted here is that the putative simple cubic perovskite CaCoO$_{3}$ has been predicted to be FM and metallic as well as SrCoO$_{3}$.\cite{17} On the other hand, the crystal structure of CaCoO$_{3}$ is expected to be similar to that of the GdFeO$_{3}$-type perovskite CaFeO$_{3}$, since the radius of Co$^{4+}$ ion (0.53 \AA) is close to that of Fe$^{4+}$ ion (0.58 \AA). Therefore, it is of great interest how the structure and the electronic ground state of the Co$^{4+}$-containing perovskite evolve as the A-site ion changes from Sr to Ca. In this letter, we report high-pressure synthesis, crystal structure, and electronic properties of a new perovskite oxide CaCoO$_{3}$. Unexpectedly, CaCoO$_{3}$ is found to be antiferromagnetic (AFM) below 95 K. From the structural viewpoint, CaCoO$_{3}$ is similar to CaFeO$_{3}$ with an orthorhombic distortion. However, unlike the case of CaFeO$_{3}$, the electronic ground state of CaCoO$_{3}$ is found to be nearly metallic and free from the charge disproportionation at low temperatures. We will discuss the origin of the AFM ordering, which is potentially HM, and the reason for the absence of the charge disproportionation in CaCoO$_{3}$. 

Polycrystalline sample of CaCoO$_{3}$ was prepared by a solid state reaction under high pressure in an oxidizing atmosphere. The synthesis procedure for CaCoO$_{3}$ is as follows. The starting materials, CaCO$_{3}$ and Co$_{3}$O$_{4}$, were stoichiometrically mixed. The mixture was heated at 1173 K for 12 h in air. The obtained powder was pelletized and sintered again at 1373 K for 24 h in a flow of oxygen (1 atm), followed by quenching into water at room temperature. Through these processes, an oxygen-deficient perovskite (Brownmillerite-like) Ca$_{2}$Co$_{2}$O$_{5+\delta}$ was synthesized. The quenched pellet was then pulverized and packed into a gold capsule together with an oxidizer NaClO$_{3}$. The capsule was annealed at 753 K for 1 h at a high pressure of 8 GPa using a cubic-anvil-type apparatus. The synchrotron powder X-ray diffraction (XRD) with an wavelength of 0.68975 $\AA$ was carried out at BL-8B, Photon Factory, KEK, Japan. The diffraction patterns were analyzed by the Rietveld refinement using RIETAN-FP.\cite{18} Thermogravimetric (TG) analysis was performed on a Netzsch TG-DTA2500-IW thermal analyzer in an atmosphere of 96 \% Ar and 4 \% H$_{2}$ with a heating rate of 15 ${}^\circ$C/min. The magnetization $M$, electrical resistivity $\rho$, and specific heat $C$ were measured using MPMS and PPMS manufactured by Quantum Design, respectively.   

\begin{figure}[t]
\begin{center}
\includegraphics[width=8cm]{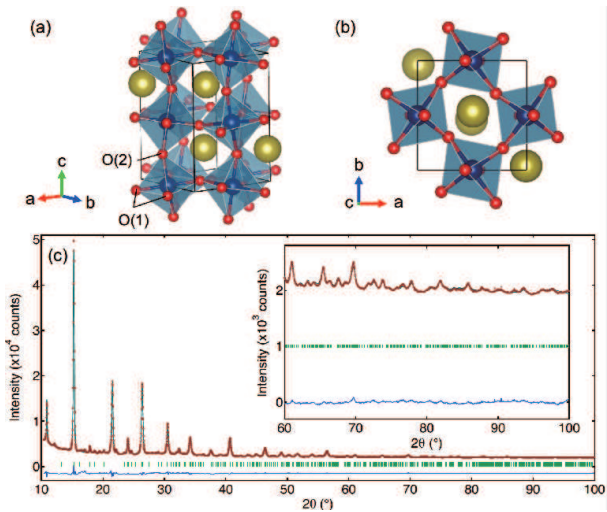}
\caption{(color online) (a) Crystal structure of CaCoO$_{3}$ and (b) the projection along the $c$ axis. The blue and yellow spheres correspond to Co and Ca ions, respectively. (c) Observed (red circles), calculated (green line), and differential (blue line) synchrotron powder XRD patterns. The green tick marks indicate the calculated peak positions.}
\end{center}
\end{figure}

The powder XRD pattern of CaCoO$_{3}$ was indexed with an orthorhombic unit cell [$a = $5.2711(7) \AA, $b = $5.2954(2) \AA, and $c =$ 7.4394(4) \AA]. The Rietveld analysis reveals that the diffraction peaks are well fitted by the GdFeO$_{3}$-type structure with space group $Pbnm$ (No.62) as shown in Fig. 1(c). The crystallographic data and the reliability factors are summarized in Table I. The occupancies of all atoms were fixed to 1 during the refinements (Note that the refinement of the occupancies gives negligible improvement in the Rietveld analysis). Figures 1(a) and 1(b) show the refined structure of CaCoO$_{3}$. As seen in Table I\hspace{-.1em}I  summarizing the bond lengths and angles, the CoO$_{6}$ octahedra are slightly distorted with Co-O bond lengths in the range 1.84-1.98 $\AA$ and O-Co-O bond angles of 90$\pm$1.5$^\circ$. On the other hand, the CoO$_{6}$ octahedra are largely tilted as confirmed by the Co-O-Co bond angles of 155.6-159.5$^\circ$, which is comparable to those for CaFeO$_{3}$ (157.9-158.6$^\circ$).\cite{9} The large tilting of the CoO$_{6}$ octahedra in CaCoO$_{3}$ reflects the significant decrease of the tolerance factor $t$ from unity ($t =$0.923 and 0.898 for CaCoO$_{3}$ and CaFeO$_{3}$, respectively). Using the Ca-O and Co-O bond lengths, we estimated the bond-valence sums for Ca and Co ions,\cite{24} which are +2.17 and +4.01, respectively (Table I\hspace{-.1em}I\hspace{-.1em}I). This result indicates that the oxidation state can be described as Ca$^{2+}$Co$^{4+}$O$_{3}$.

\begin{table}[tb]
\caption{Structural parameters for CaCoO$_{3}$. Space group; $Pbnm$ (No. 62). $a = $5.2711(7) \AA, $b = $5.2954(2) \AA, $c =$ 7.4394(4) \AA, and $V=207.7$ \AA$^{3}$. $R_{\rm wp}$=2.40, $S=$1.28.}
{\renewcommand\arraystretch{1.3}
\begin{tabular}{cccccc}
\hline 
    atoms & site & $x$ & $y$ & $z$ & $B$ (\AA$^{2}$) \\
\hline
    Ca & 4c & 0.494(2) & 0.4628(5) & 0.25 & 0.58(4) \\
    Co & 4b & 0.5 & 0 & 0 & 0.33(2) \\
    O(1) & 8d & 0.783(2) & 0.199(2) & 0.034(1) & 0.90(8)\\ 
    O(2) & 4c & 0.064(3) & 0.499(1) & 0.25 & 0.90(8)\\
\hline 
\end{tabular}
}
\end{table}

\begin{table}[tb]
\caption{Evaluated bond lengths, bond angles of CaCoO$_{3}$.}
{\renewcommand\arraystretch{1.3}
 \begin{tabular}{ccccc}\hline
    bond & lengths (\AA) &\ & bonds & angles($^\circ$) \\ \hline
    Co-O(1) & 1.84(1) $\times$2 &\  \ & Co-O(1)-Co & 155.6(5) \\
     & 1.98(1) $\times$2 &\  \ & Co-O(2)-Co & 159.5(9) \\
    Co-O(2) & 1.890(3) $\times$2 &\ \  \ & O(1)-Co-O(1) & 88.6(2) $\times$2 \\
    Ca-O(1)& 2.351(9) $\times$2&\  \ \ & & 91.4(2) $\times$2\\
      & 2.54(1) $\times$2&\  \ \ & O(1)-Co-O(2) &88.6(5) $\times$2\\
      & 2.61(1) $\times$2&\  \ \ &  &89.5(7) $\times$2\\
     Ca-O(2) & 2.28(2) &\ \ \  & & 90.5(7) $\times$2\\
      &2.47(1)&\  \ \ & & 91.4(5) $\times$2\\\hline
  \end{tabular}
}
\end{table}

 \begin{table}[htb]
\caption{Bond-valence sums (BVS) of CaCoO$_{3}$. BVS are calculated by $\sum_i \mathrm{exp}\{(r_0-r_i)/0.37\}$, where $r_0=1.967$ for Ca$^{2+}$ and $r_0=1.75$ for Co$^{4+}$.}
{\renewcommand\arraystretch{1.3}
  \begin{tabular}{ccccc}\hline
    \quad atoms \qquad& \quad Ca &\quad Co &\quad O(1) &\quad O(2) \\ \hline
    \quad BVS \qquad &\quad+2.17\qquad &\quad+4.01\qquad&\quad-2.06\qquad &\quad-2.06\qquad\\
    \hline
  \end{tabular}
}
\end{table}

TG curve for CaCoO$_{3}$ is shown in Fig. 2. The significant weight loss owing to the release of oxygen is observed in several steps. Since the final product was identified to be a mixture of CaO and Co by XRD experiments, the decomposition can be expressed as,
\begin{equation}
\begin{aligned}
\rm{CaCoO_{3}} \longrightarrow \rm{CaO}+\rm{Co}+\rm{O_{2}}\uparrow.
\end{aligned}
\label{eq:ep1}
\end{equation}                 
The observed weight loss of 22.3 \% in the temperature range about 700 ${}^\circ$C is almost the same as or even slightly larger than the ideal value of 21.8 \% for stoichiometric CaCoO$_{3}$, indicating that our sample is free from oxygen deficiency. The weight loss at around 180 ${}^\circ$C and 300 ${}^\circ$C can be associated with structural changes to oxygen-deficient-ordered perovskite CaCoO$_{3-\delta}$ ($\delta \sim 0.5$), which adopts presumably the Brownmillerite-type structure.

\begin{figure}[t]
\begin{center}
\includegraphics[width=8cm]{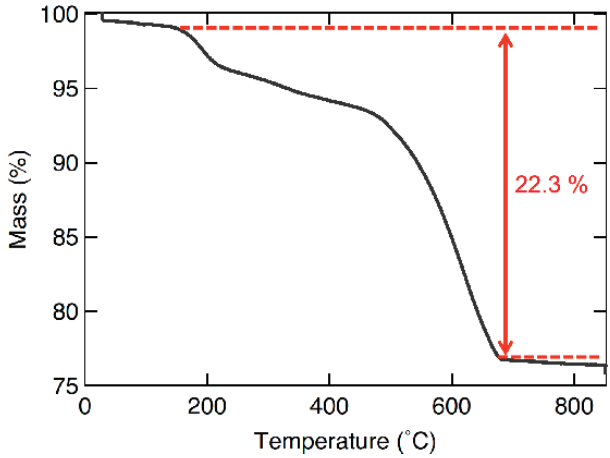}
\caption{(color online). Thermogravimetry (TG) for CaCoO$_{3}$ measured on heating in 96 \% Ar-4 \% H$_{2}$ atmosphere.}
\end{center}
\end{figure}

Figure 3(a) shows the temperature dependence of the magnetic susceptibility measured under a magnetic field of 0.1 T in zero-field cooling (ZFC) and field cooling (FC) runs. An abrupt drop at 95 K ($=T_{\rm N}$) indicates the onset of an antiferromagnetic (AFM) order, which is followed by anomalies at lower temperatures, presumably corresponding to the spin reorientations. The data between 200 and 300 K were fit to the Curie-Weiss law, $\chi(T) = C/(T-\theta)$ as shown in the form of the inverse susceptibility plotted in the inset of Fig. 3(a). Here, $C$ is the Curie constant and $\theta$ is the Weiss temperature, which are estimated as $C =$ 1.67 emu K/mol and $\theta =$ 91.7 K, respectively. The effective Bohr magnetons number $P_{\rm eff} = 3.65$ $\rm \mu _{\rm B}$ estimated from the Curie constant is close to the theoretical value of 3.87 $\rm \mu _{\rm B}$ expected for the intermediate spin state of Co$^{4+}$ ($e_{\rm g}^{1}t_{\rm 2g}^{4}$) with $S =$ 3/2. Thus, it is likely that the spin state of CaCoO$_{3}$ is in the intermediate spin state as well as SrCoO$_{3}$, whereas the ordered spin structure of CaCoO$_{3}$ is apparently AF in contrast to the FM state in SrCoO$_{3}$. 

The temperature dependence of $\rho$ is shown in Fig. 3 (b). As compared to the metallic SrCoO$_{3}$, CaCoO$_{3}$ is seemingly nonmetallic. Despite the nonmetallic behavior, the magnitude of $\rho$ at room temperature is as low as 3 m$\Omega$cm and $\rho$ increases only slightly upon cooling. Such behavior is often seen in strongly correlated transition-metal oxides near the Mott insulating state, which is called an incoherent "bad metal''. Thus, CaCoO$_{3}$ is considered to be an incoherent metallic state. At around $T_{\rm N}$, a small kink corresponding to the AFM transition is observed, which is clearly discernible in the derivative of the resistivity $d\rho/dT$. The point to note here is that a negative magnetoresistance ($\Delta\rho/\rho$(0 T)$=[\rho$(9 T$)-\rho$(0 T$)]/\rho$(0 T)) is seen around $T_{\rm N}$ as shown in the inset of Fig. 3(b). Such a negative magnetoresistance is irrelevant to a simple G-type AFM order (all the neighboring spins are aligned oppositely) and is typically found around the FM or HM critical temperature, where the spin fluctuation tends to be suppressed by the magnetic fields of several Teslas.

To characterize the nearly metallic state of CaCoO$_{3}$, we measured the temperature dependence of specific heat as shown in the form of $C/T$ versus $T^{2}$ (see Fig. 4). The solid line is the fitting with the equation $C/T=\gamma + (12/5)\pi ^{4}NR\Theta _{\rm D}^{3}T^{2}$ ($R=8.31$ J/mol K and $N=4$), where $\gamma$ and $\Theta _{\rm D}$ represent an electronic specific heat coefficient and Debye temperature, respectively. Thus we obtained the values of $\gamma=33.0$ mJ/mol K$^{2}$ and $\Theta _{\rm D}=450$ K. The electronic specific heat of CaCoO$_{3}$ is much larger than that of the conventional metals, implying that the electron correlation enhances the effective mass of conduction electrons. On the other hand, the electronic specific heat of CaCoO$_{3}$ is smaller than that of metallic and cubic SrCoO$_{3}$ ($\sim$ 45.7 mJ/mol K$^{2}$),\cite{21} which signifies that CaCoO$_{3}$ has a pseudo gap near the Fermi level. The formation of the pseudo gap in CaCoO$_{3}$ is consistent with the expectation that the system is situated close to the Mott insulating state upon the band-width narrowing through the lattice distortion. This conjecture is also compatible with the nonmetallic behavior found in the $T$ dependence of $\rho$. However, since the grain boundary scattering affects on the $T$ dependence of $\rho$, detailed studies on the single crystalline sample are desirable to clarify the origin of the nonmetallic behavior in CaCoO$_{3}$.

\begin{figure}[t]
\begin{center}
\includegraphics[width=8cm]{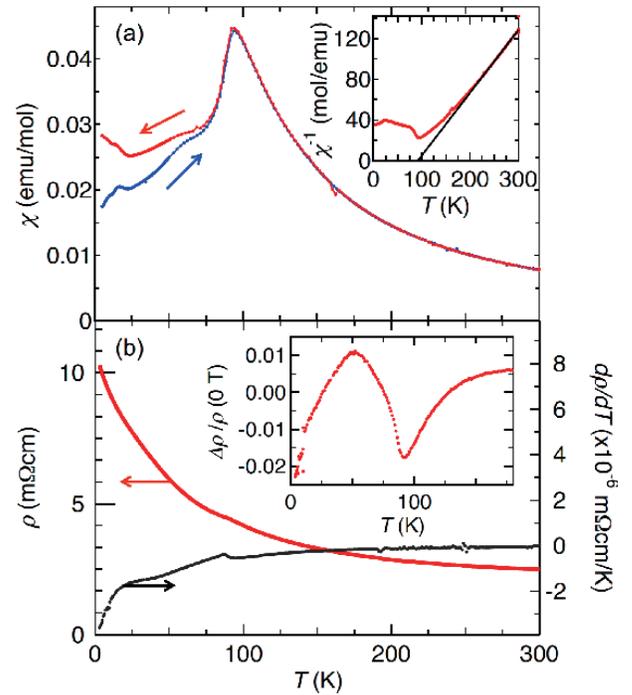}
\caption{(color online). (a) Temperature dependence of the magnetic susceptibility $\chi$ in an applied magnetic field of 0.1 T measured on heating after zero-field cooling (blue) and on field cooling (red) for CaCoO$_{3}$. The arrows indicate the direction of temperature change. Inset shows the inverse magnetic susceptibility 1/$\chi$ as a function of temperature. (b) Temperature dependence of resistivity $\rho$ (left axis) and the temperature derivative of resistivity $d\rho/dT$ (right axis). Inset shows the magnetoresistance ($\Delta\rho/\rho$(0 T)$=[\rho$(9 T$)-\rho$(0 T$)]/\rho$(0 T)) as a function of temperature.}
\end{center}
\end{figure}

\begin{figure}[t]
\begin{center}
\includegraphics[width=8cm]{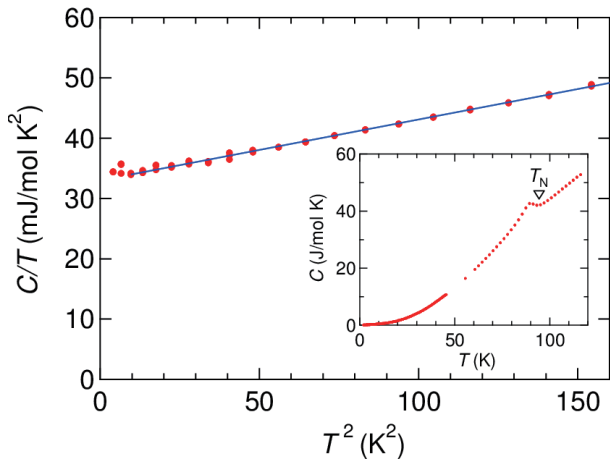}
\caption{(color online). $T^{2}$ dependence of $C/T$ for CaCoO$_{3}$. The blue solid line shows the fitting of the experimental data with the equation $C/T=\gamma +\beta T^{2}$. The inset shows $C$ as a function of temperature, where an anomaly corresponding to $T_{\rm N}$ is observed.}
\end{center}
\end{figure}

\begin{figure}[t]
\begin{center}
\includegraphics[width=8cm]{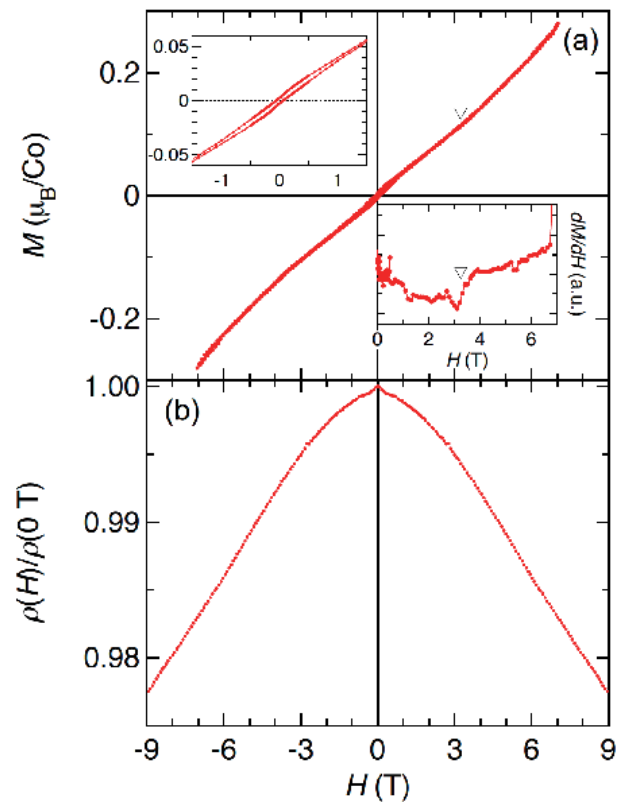}
\caption{(color online). Magnetic field dependence of (a) the magnetization, and (b) the resistivity at 10 K. Upper and lower insets in (a) are the magnetization in the low field region and the field derivative of the magnetization $dM/dH$, respectively.}
\end{center}
\end{figure}

Considering the previous reports that the ground state of Sr$_{1-x}$Ca$_{x}$CoO$_{3}$ remains FM with $x$ up to 0.8,\cite{21} and the FM metallic phase with GdFeO$_{3}$-type structure emerges in La$_{1-x}$Ca$_{x}$CoO$_{3}$ with $x$ above 0.1,\cite{3,19} the observation of the AFM transition in CaCoO$_{3}$ is unexpected. To further characterize the magnetic ground state of CaCoO$_{3}$, we measured the field dependence of $M$ at 10 K as shown in Fig. 5(a). The $M-H$ curve shows a small hysteretic behavior without reaching saturation even at 7 T, indicating that the magnetic ground state is AFM with a tiny FM component. As shown in the derivative of the magnetization $dM/dH$, an anomaly is found at 3.5 T, presumably corresponding to an $H$-induced spin reorientation. This behavior implies that the ground state is not a simple G-type AFM state. Here, we presume that the AFM phase found in CaCoO$_{3}$ has HM spin structure as in the case of SrFeO$_{3}$. This is supported by the facts that $\theta$ is positive and its magnitude is comparable to $T_{\rm N}$, which signifies that the neighboring spins in the ground state are aligned nearly parallel. The negative magnetoresistance discernible at low temperatures (see Fig. 5(b)) implies the gradual change in the spin structure to the FM state.

For the origin of the HM order in oxides with Fe$^{4+}$ ions, two possible mechanisms have been considered: (i) the double exchange model suitable for describing the magnetic ground state of the unusually high valence system with the negative $p-d$ charge-transfer energy $\it \Delta$ \cite{7,22}; (ii) the competition between the FM double exchange interaction and the AFM superexchange interaction.\cite{29,30} Since SrCoO$_{3}$ has negative $\it \Delta \sim$ -2 eV (Ref.\cite{14}), it is presumable that the double exchange mechanism for the transition-metal oxides with negative $\it \Delta$ works also in CaCoO$_{3}$. If this is the case, considering that the magnetic ground state in this model can be either FM or HM depending on the strength of the $p-d$ hybridization, CaCoO$_{3}$ with the orthorhombic distortion can adopt the HM state instead of the FM state appeared in the cubic perovskite SrCoO$_{3}$. On the other hand, given that the FM double exchange interaction works in SrCoO$_{3}$, it is also presumable that the introduction of the orthorhombic distortion by the Ca substitution for Sr changes the ground state from the FM to the HM state, where the FM double exchange interaction is so reduced as to be competitive with the AFM superexchange interaction. In either case, the HM state likely appears in CaCoO$_{3}$ as a consequence of the band-width narrowing due to the GdFeO$_{3}$-type orthorhombic distortion.


Finally, let us discuss the difference in the electronic ground state between CaCoO$_{3}$ and other distorted perovskites containing unusually high valence Fe and Ni ions, typified by CaFeO$_{3}$. The charge disproportionation is ubiquitously found in distorted perovskites with Fe$^{4+}$ \cite{26,27,31,32} and Ni$^{3+}$.\cite{33,34} The origin of the charge disproportionation has been discussed in terms of the enhanced electron-phonon interaction coupled to the breathing mode\cite{10} and the Hund's coupling with strong $pd$ hybridization.\cite{11,12}  The point is that the doubly degenerate $e_{\rm g}$ orbitals occupied by one electron are at the Fermi level in these perovskites, which is suitable for the breathing-type lattice distortion. On the other hand, for CaCoO$_{3}$ with the intermediate spin state, not only the $e_{\rm g}$ orbitals but the $t_{\rm 2g}$ orbitals are located near the Fermi level. Therefore, the absence of the charge disproportionation in CaCoO$_{3}$ signifies that the electronic instability toward the localized state in the largely distorted lattice with the narrow band width is avoided by the presence of the itinerant $t_{\rm 2g}$ electrons, as suggested by the first-principles calculations for SrCoO$_{3}$ \cite{16,17}. 


To summarize, we have succeeded in synthesizing the GdFeO$_{3}$-type perovskite CaCoO$_{3}$ without oxygen deficiency. The bond-valence sum calculations based on the refined crystal structure show that the valences of Ca and Co are +2 and +4, respectively. The magnetization and resistivity measurements indicate that CaCoO$_{3}$ is an antiferromagnetic and incoherent metal. The positive Weiss temperature and the nonlinear magnetization curve at low temperature imply that the magnetic structure is noncollinear, potentially helimagnetic. 
This work demonstrates the rich magnetism in perovskite oxides with Co$^{4+}$ ions and the significant effect of the band-width narrowing on the magnetic ground state in (Sr,Ca)CoO$_{3}$.


\section{ACKNOWLEDGMENTS}
The authors appreciate M. Takano, H. Sakai and J. Fujioka for the helpful suggestions.
This work is partly supported by JSPS, KAKENHI (Grants No. 17H01195 and No. 16K17736), JST PRESTO Hyper-nano-space design toward Innovative Functionality (JPMJPR1412), Asahi Grass Foundation, and Grant for Basic Science Research Projects of the Sumitomo Foundation. The powder XRD measurement was performed with the approval of the Photon Factory Program Advisory Committee (Proposal No. 2015S2-007).

\end{document}